
\newif\ifTextures
\Texturesfalse

\ifTextures
\input epsf.def
\else
\input epsf.tex
\fi

%

\def\unlock{
 \catcode`@=11 }
\unlock

 \font\fourteenrm=cmr10                 scaled\magstep2
 \font\twelverm=cmr10                   scaled\magstep1
 \font\elevenrm=cmr10                   scaled 1095
 \font\ninerm=cmr9
 \font\sixrm=cmr6

 \font\fourteenbf=cmbx10                scaled\magstep2
 \font\twelvebf=cmbx10                  scaled\magstep1
 \font\elevenbf=cmbx10                  scaled 1095
 \font\ninebf=cmbx9
 \font\sixbf=cmbx5
 \font\twentyfouri=cmmi10 scaled\magstep4  \skewchar\twentyfouri='177
 \font\seventeeni=cmmi10  scaled\magstep3  \skewchar\seventeeni='177
 \font\fourteeni=cmmi10   scaled\magstep2  \skewchar\fourteeni='177
 \font\twelvei=cmmi10     scaled\magstep1  \skewchar\twelvei='177
 \font\eleveni=cmmi10     scaled 1095      \skewchar\eleveni='177
 \font\ninei=cmmi9                         \skewchar\ninei='177
 \font\sixi=cmmi6                          \skewchar\sixi='177
 \font\twentyfoursy=cmsy10 scaled\magstep4 \skewchar\twentyfoursy='60
 \font\seventeensy=cmsy10  scaled\magstep3 \skewchar\seventeensy='60
 \font\fourteensy=cmsy10   scaled\magstep2 \skewchar\fourteensy='60
 \font\twelvesy=cmsy10     scaled\magstep1 \skewchar\twelvesy='60
 \font\elevensy=cmsy10     scaled 1095     \skewchar\elevensy='60
 \font\ninesy=cmsy9                        \skewchar\ninesy='60
 \font\sixsy=cmsy6                         \skewchar\sixsy='60

 \font\fourteenex=cmex10    scaled\magstep2
 \font\twelveex=cmex10      scaled\magstep1
 \font\elevenex=cmex10      scaled 1095
 \font\tenex=cmex10
 \font\nineex=cmex10 at 9pt

 \font\fourteensl=cmsl10    scaled\magstep2
 \font\twelvesl=cmsl10      scaled\magstep1
 \font\elevensl=cmsl10      scaled 1095
 \font\ninesl=cmsl10 at 9pt

 \font\fourteenit=cmti10   scaled\magstep2
 \font\twelveit=cmti10     scaled\magstep1
 \font\elevenit=cmti10     scaled 1095
 \font\tenit=cmti10
 \font\nineit=cmti10 at 9pt
 
 \font\twelvett=cmtt10     scaled\magstep1
 \font\eleventt=cmtt10     scaled 1095
 \font\tentt=cmtt10
 \font\ninett=cmtt10 at 9pt
 
 \font\twelvecp=cmcsc10    scaled\magstep1
 \font\elevencp=cmcsc10    scaled 1095
 \font\tencp=cmcsc10
 \font\ninecp=cmcsc10 at 9pt

 \newfam\cpfam
 \newcount\f@ntkey            \f@ntkey=0
 \def\samef@nt{\relax \ifcase\f@ntkey \rm \or\oldstyle \or\or
          \or\it \or\sl \or\bf \or\tt \or\caps \fi }

\def\fourteenpoint{\relax
  \textfont0=\fourteenrm \scriptfont0=\tenrm \scriptscriptfont0=\sevenrm
   \def\rm{\fam0 \fourteenrm \f@ntkey=0 }\relax
  \textfont1=\fourteeni \scriptfont1=\teni \scriptscriptfont1=\seveni
   \def\oldstyle{\fam1 \fourteeni\f@ntkey=1 }\relax
  \textfont2=\fourteensy \scriptfont2=\tensy \scriptscriptfont2=\sevensy
  \textfont3=\fourteenex \scriptfont3=\twelveex \scriptscriptfont3=\tenex
   \def\it{\fam\itfam \fourteenit\f@ntkey=4 }
  \textfont\itfam=\fourteenit
   \def\sl{\fam\slfam \fourteensl\f@ntkey=5 }
  \textfont\slfam=\fourteensl \scriptfont\slfam=\tensl
   \def\bf{\fam\bffam \fourteenbf\f@ntkey=6 }
  \textfont\bffam=\fourteenbf \scriptfont\bffam=\tenbf
  \scriptscriptfont\bffam=\sevenbf
   \def\tt{\fam\ttfam \twelvett  \f@ntkey=7 }
  \textfont\ttfam=\twelvett
   \def\caps{\fam\cpfam \twelvecp\f@ntkey=8 }
  \textfont\cpfam=\twelvecp
  \samef@nt }

\def\twelvepoint{\relax
  \textfont0=\twelverm \scriptfont0=\ninerm \scriptscriptfont0=\sixrm
   \def\rm{\fam0 \twelverm \f@ntkey=0 }\relax
  \textfont1=\twelvei \scriptfont1=\ninei \scriptscriptfont1=\sixi
   \def\oldstyle{\fam1 \twelvei\f@ntkey=1 }\relax
  \textfont2=\twelvesy \scriptfont2=\ninesy \scriptscriptfont2=\sixsy
  \textfont3=\twelveex \scriptfont3=\elevenex \scriptscriptfont3=\tenex
   \def\it{\fam\itfam \twelveit \f@ntkey=4 }
  \textfont\itfam=\twelveit
   \def\sl{\fam\slfam \twelvesl \f@ntkey=5 }
  \textfont\slfam=\twelvesl \scriptfont\slfam=\ninesl
   \def\bf{\fam\bffam \twelvebf \f@ntkey=6 }
  \textfont\bffam=\twelvebf \scriptfont\bffam=\ninebf
  \scriptscriptfont\bffam=\sixbf
   \def\tt{\fam\ttfam \twelvett \f@ntkey=7 }
  \textfont\ttfam=\twelvett
   \def\caps{\fam\cpfam \twelvecp \f@ntkey=8 }
  \textfont\cpfam=\twelvecp
  \samef@nt }
 
\def\elevenpoint{\relax
  \textfont0=\elevenrm \scriptfont0=\ninerm \scriptscriptfont0=\sixrm
   \def\rm{\fam0 \elevenrm \f@ntkey=0 }\relax
  \textfont1=\eleveni \scriptfont1=\ninei \scriptscriptfont1=\sixi
   \def\oldstyle{\fam1 \eleveni\f@ntkey=1 }\relax
  \textfont2=\elevensy \scriptfont2=\ninesy \scriptscriptfont2=\sixsy
  \textfont3=\elevenex \scriptfont3=\tenex \scriptscriptfont3=\tenex
   \def\it{\fam\itfam \elevenit \f@ntkey=4 }
  \textfont\itfam=\elevenit
   \def\sl{\fam\slfam \televensl \f@ntkey=5 }
  \textfont\slfam=\elevensl \scriptfont\slfam=\ninesl
   \def\bf{\fam\bffam \elevenbf \f@ntkey=6 }
  \textfont\bffam=\elevenbf \scriptfont\bffam=\ninebf
  \scriptscriptfont\bffam=\sixbf
   \def\tt{\fam\ttfam \eleventt \f@ntkey=7 }
  \textfont\ttfam=\eleventt
   \def\caps{\fam\cpfam \elevencp \f@ntkey=8 }
  \textfont\cpfam=\elevencp
  \samef@nt }

 \def\tenpoint{\relax
  \textfont0=\tenrm \scriptfont0=\sevenrm \scriptscriptfont0=\fiverm
   \def\rm{\fam0 \tenrm \f@ntkey=0 }\relax
  \textfont1=\teni \scriptfont1=\seveni \scriptscriptfont1=\fivei
   \def\oldstyle{\fam1 \teni \f@ntkey=1 }\relax
  \textfont2=\tensy \scriptfont2=\sevensy \scriptscriptfont2=\fivesy
  \textfont3=\tenex \scriptfont3=\tenex \scriptscriptfont3=\tenex
   \def\it{\fam\itfam \tenit \f@ntkey=4 }
  \textfont\itfam=\tenit
   \def\sl{\fam\slfam \tensl \f@ntkey=5 }
  \textfont\slfam=\tensl
   \def\bf{\fam\bffam \tenbf \f@ntkey=6 }
  \textfont\bffam=\tenbf \scriptfont\bffam=\sevenbf
  \scriptscriptfont\bffam=\fivebf
   \def\tt{\fam\ttfam \tentt \f@ntkey=7 }
  \textfont\ttfam=\tentt
   \def\caps{\fam\cpfam \tencp \f@ntkey=8 }
  \textfont\cpfam=\tencp
  \setbox\strutbox=\hbox{\vrule height 8.5pt depth 3.5pt width\z@}
  \samef@nt }

\def\ninepoint{\relax
  \textfont0=\ninerm \scriptfont0=\sevenrm \scriptscriptfont0=\fiverm
   \def\rm{\fam0 \ninerm \f@ntkey=0 }\relax
  \textfont1=\ninei \scriptfont1=\seveni \scriptscriptfont1=\fivei
   \def\oldstyle{\fam1 \ninei \f@ntkey=1 }\relax
  \textfont2=\ninesy \scriptfont2=\sevensy \scriptscriptfont2=\fivesy
  \textfont3=\nineex \scriptfont3=\nineex \scriptscriptfont3=\nineex
   \def\it{\fam\itfam \nineit \f@ntkey=4 }
  \textfont\itfam=\nineit
   \def\sl{\fam\slfam \ninesl \f@ntkey=5 }
  \textfont\slfam=\ninesl
   \def\bf{\fam\bffam \ninebf \f@ntkey=6 }
  \textfont\bffam=\ninebf \scriptfont\bffam=\sevenbf
  \scriptscriptfont\bffam=\fivebf
   \def\tt{\fam\ttfam \ninett  \f@ntkey=7 }
  \textfont\ttfam=\ninett
   \def\caps{\fam\cpfam \ninecp\f@ntkey=8 }
  \textfont\cpfam=\ninecp
  \setbox\strutbox=\hbox{\vrule height 8.5pt depth 3.5pt width\z@}
  \samef@nt }

%
\nopagenumbers
\headline={\rm\hss\folio\hss}

\def\abstract#1{ \setbox0=\vbox{\hsize=112mm\tenpoint
\baselineskip=11pt \parindent=17pt
\vskip -\baselineskip \vskip 8mm
 #1
\vglue -\baselineskip \vglue 6.5mm
\noindent PACS numbers:
\vglue -1mm}
\line{\hfil\box0\hfil} }

\newcount\sectionnumber    \sectionnumber=0
\newcount\appendixnumber   \appendixnumber=0
\newtoks\appendixstyle     \global\appendixstyle={\Alphabetic}

\def\section#1{\goodbreak\vskip 0.65cm
 \global\advance\sectionnumber by 1
 \line{\hfil\bf\the\sectionnumber . #1\hfil}
 \vskip 0.45cm }

\def\apxn{\the\appendixstyle{\the\appendixnumber}}
\def\Alphabetic#1{\count255=64 \advance\count255 by #1\char\count255}
\def\appendix#1{\goodbreak\vskip 0.65cm
\global\advance\appendixnumber by 1
\centerline{\tenpoint APPENDIX \apxn.}
\vskip 0.45cm
\centerline{\bf #1}
\vskip 0.45cm}

\def\references{\goodbreak\vskip 0.65cm\tenpoint
 \centerline{REFERENCES}
 \vskip 0.45cm }

\hsize=126truemm
\vsize=190truemm
\parindent=19.2pt
\baselineskip=12.7pt plus 0.2pt
\elevenpoint

\def\today{December 4, 1997}



\hyphenation{de-ri-va-tive}
\hyphenation{renor-ma-li-za-ble}
\hyphenation{renor-ma-li-za-tion}
\hyphenation{renor-ma-li-zed}
\hyphenation{ano-ma-lous}
\hyphenation{fer-mion}

\def\smn#1#2{\hbox{${#1 \over #2}$}}

%
%
%

\hfill\today\vskip -2.7mm  

\vglue 20mm
\hyphenation{com-po-nent}
\hyphenation{na-me-ly}
\twelvepoint \baselineskip=16pt
\centerline{TWO COMPONENT THEORY AND}
\centerline{ELECTRON MAGNETIC MOMENT}
\ninepoint \baselineskip=11pt
\vskip -\baselineskip \vskip 8.5mm
\centerline{\elevenpoint\caps M. Veltman}
\vskip -\baselineskip \vskip 8mm
\centerline{Department of Physics, University of Michigan}
\centerline{Ann Arbor, MI 48109, USA}
\centerline{and}
\centerline{CERN, Geneva, Switzerland}
\vskip -\baselineskip \vskip 8.8mm

\tenpoint
\centerline{\it (Received \phantom{June 27, 1991})}

\abstract{The two-component formulation of quantum electrodynamics is studied.
The relation with the usual Dirac formulation is exhibited, and the Feynman
rules for the two-component form of the theory are presented in terms of
familiar objects. The transformation from the Dirac theory to the
two-component theory is quite amusing, involving Faddeev-Popov ghost loops
of a fermion type with bose statistics. The introduction of an anomalous
magnetic moment in the two-component formalism is simple; it is not
equivalent to a Pauli term in the Dirac formulation. Such an anomalous magnetic
moment appears not to destroy the renormalizability of the theory but violates
unitarity.}

\baselineskip=12.7pt plus 0.2pt
\elevenpoint

\section{Introduction}%
In 1958, Feynman and Gell-Mann [1] revived the two-component fermion theory
in the context of their work on the V-A form of the weak interactions.
This was studied extensively by several authors [2-4], who reformulated quantum
electrodynamics in the form of a two-component theory, and also applied the
formalism to the then current theory of weak interactions, the V-A theory.
While nobody really followed up on this development, there are nonetheless
certain advantages to the two-component theory, as it shows quite clearly
and separately the electric and magnetic interactions of the electron with
the photon field.

We review the two-component formulation in a somewhat more familiar form, that
is in terms of ordinary Dirac spinors. This also exhibits very clearly
the relationship between the Dirac form and the two-component form of the
theory.

In the case of Compton scattering one may compute the cross-section
using that two-component formalism, in the expectation that it then becomes
clear which part of the cross-section is due to the magnetic moment of
the electron. This indeed can be done. Varying the electron magnetic moment
(i.e. adding an anomalous part) in the two-component formalism can be done
quite easily; one might think that this amounts to the addition of a Pauli
term in the Dirac formulation of the theory, but this is not the case. Some
remarks to this effect have been made already by Brown [2]. The addition
of an anomalous magnetic moment in the two-component theory appears not to
destroy the property of renormalizability, but on the other hand unitarity
is no longer maintained. Translating the so introduced moment into the Dirac
form one finds a complicated non-polynomial interaction; in lowest order it
gives the same contribution as the Pauli term.

We use the Pauli metric, i.e. $g_{\mu\nu}$ is the unit matrix.
For conventions concerning spinors, gamma matrices, etc. see Ref. [5].

\section{The two-component Lagrangian}%
The Lagrangian of quantum electrodynamics (QED) in the Dirac formulation is
$$ {\cal L}^{qed}_{\rm D} = - \smn12 \partial_\mu A_\nu \partial_\mu A_\nu
 - \overline \psi (\gamma^\mu D_\mu + m)\psi\, ,$$
with as usual $D_\mu = \partial_\mu + ieA_\mu$. We now substitute
$$ \psi \rightarrow (- \gamma^\nu D_\nu + m) \psi$$
without changing $\overline \psi$. Thus we obtain
$$\eqalignno{{\cal L}^{\rm QED}_2 = &- \smn12 \partial_\mu A_\nu \partial_\mu A_\nu\cr
 &- \overline \psi \left( - \partial^2
 - ie\gamma^\mu\gamma^\nu \partial_\mu A_\nu - 2ie A_\nu\partial_\nu
 + e^2A_\mu A_\mu + m^2\right)\psi\cr
 = &- \smn12 \partial_\mu A_\nu \partial_\mu A_\nu\cr
 &- \overline\psi \left(- \partial^2
 - 2ie \sigma^{\mu\nu} \partial_\mu A_\nu -ie \partial_\mu A_\mu
 -2ieA_\nu\partial_\nu + e^2 A^2 + m^2\right)\psi\cr
 = &- \smn12 \partial_\mu A_\nu \partial_\mu A_\nu\cr
 &-(\partial_\mu -ieA_\mu) \overline \psi (\partial_\mu +ie A_\mu)\psi
 - m^2 \overline \psi \psi
 + 2ie\left( \overline \psi \sigma^{\mu\nu} \psi\right) \partial_\mu A_\nu\, .}$$
Here $\sigma^{\mu\nu} = \smn14 (\gamma^\mu\gamma^\nu - \gamma^\nu\gamma^\mu)$.
Except for the last term this Lagrangian is precisely that of a charged scalar
particle of mass $m$ interacting with the e.m. field in the standard way; the last
term, containing $\sigma^{\mu\nu}$, is the magnetic part of the interaction. Since
in this Lagrangian any term has either no gamma matrix or two, one may
rewrite the above in terms of right- and left-handed spinors. With
$$\eqalignno{\psi_L &= \smn{1 + \gamma^5}2\,\psi,\quad\quad
\overline\psi_L = \overline\psi\,\smn{1 - \gamma^5}2\cr
\noalign{\noindent\rm and}
\quad \psi_R &= \smn{1 - \gamma^5}2\,\psi,\quad\quad
\overline\psi_R = \overline\psi\,\smn{1 + \gamma^5}2\quad }$$
the Lagrangian will only have terms containing $\overline \psi_L \psi_R$ or
$\overline \psi_R \psi_L$.
The Lagrangian of the two-component theory is obtained by keeping only the terms
of the form $\overline \psi_R\psi_L$, thus only terms containing $1+\gamma^5$,
which makes the Lagrangian non-hermitean. That Lagrangian can be found, in
a slightly different notation, in ref. [2]. Most of the work on the two-component
theory can be viewed, from our perspective, as demonstrating that the omitted
piece of the Lagrangian (the terms of the form $\overline \psi_L\psi_R$) gives
the same contribution to the $S$-matrix as the part kept. In the two-component
formulation unitarity of the $S$-matrix is not obvious, as also emphasized
in ref. [3-4].

In this article we will use the term two-component theory in different ways.
The precise two-component theory is the theory in terms of $\overline \psi_R$
and $\psi_L$, with a non-hermitean Lagrangian. The theory that we obtained
from the Dirac theory by the transformation shown, with the Lagrangian in terms
of the familiar Dirac spinors without any $\gamma^5$, will be called the
`two-component' theory, with quotation marks. It is really a theory in terms of
four-component spinors.

The transformation shown is actually far from trivial from a field theoretical
point of view. While it may seem to be a local transformation, thus not giving
rise to a Faddeev-Popov ghost, a slightly more precise consideration shows that
the transformation involves a non-local part, and therefore Faddeev-Popov ghost
loops must be included. We will exhibit this on a purely formal level and also
directly in terms of diagrams.

\section{Faddeev-Popov ghost loops}%
It is quite instructive to consider first a very simplified theory, involving
one fermion and one scalar field with the simplest possible interaction:
$${\cal L} = {\cal L}_{\rm sc} 
 - \overline \psi(\gamma^\mu \partial_\mu + m)\psi + g(\overline \psi \psi)\varphi\,,
 \quad {\rm with}\quad{\cal L}_{\rm sc} =
 - \smn12 \varphi(-\partial^2 + m^2)\varphi\, .$$
This has a fermion propagator as usual and a very simple vertex:
$$ {\rm Propagator:}\quad {-i\gamma p + m \over p^2+m^2-i\epsilon}
 \quad\quad\quad\quad\quad\quad{\rm Vertex:}\quad g$$
Now we do the transformation $\psi \rightarrow (-\gamma^\mu \partial_\mu + m)\psi$.
The Lagrangian becomes:
$${\cal L} = {\cal L}_{\rm scalar}
 - \overline \psi(-\partial^2 + m^2)\psi
 + g \left(\overline \psi (-\gamma^\mu \partial_\mu + m)\psi\right)\varphi \, .$$
We then have the following fermion propagator and vertex:
$${\rm Propagator:}\quad {1 \over p^2+m^2-i\epsilon}
 \quad\quad\quad\quad\quad\quad{\rm Vertex:}\quad g (-i\gamma p + m)$$
where in the vertex the (incoming) momentum is that of the $\psi$ line. Obviously,
nothing could be more trivial: the numerator of the propagator has simply been
shifted to the vertex. The theory is the same as before. It looks very strange
though; the fermion appears to have a boson type propagator.

Concerning external lines, we may note that spinors corresponding to incoming
particles or outgoing anti-particles now have the energy projection operator
$-i\gamma p + m$ in front ($p$ is the momentum of the external incoming fermion,
thus flowing into the vertex). Because such spinors obey the Dirac equation that gives
just a factor $2m$. One may either keep the energy projection operator explicitly,
at the same time providing spinors with a factor $1/\sqrt{2m}$, or else keep the
spinors as usual but omitting the energy projection operator at the external lines.

Since the diagrams are literally the same as in the original theory there is no
question that the theory is still unitary after the transformation, using
at the external lines the same spinors that satisfy the Dirac equation. Can
this be seen directly?

If we consider the $S$-matrix generated by the new Feynman rules, and if we
restrict ourselves to spinors at the outgoing lines that satisfy the Dirac
equation then it is not immediately obvious that the $S$-matrix is unitary,
because a cut propagator has a residue of 1, rather then $-i\gamma p + m$.
However, this cut propagator has a factor $-i\gamma p + m$ on one side, and since
$$ -i\gamma p + m = {(-i\gamma p + m)^2 \over 2m}\, +\, {p^2 + m^2 \over 2m}\, ,$$
which is equal to $(-i\gamma p + m)^2/2m$ on mass shell,
we see that we could equally well have taken the residue $(-i\gamma p + m)/2m$
for the cut propagator. In other words, the $S$-matrix is unitary if we restrict
ourselves to external lines with spinors that obey the Dirac equation. Those
spinors must be provided with a factor $1/\sqrt{2m}$, as already noted earlier.

It is noteworthy that the new Lagrangian is not hermitean; the hermitean conjugate
differs by an interaction of the form
$$ -g\left(\overline \psi \gamma^\mu \psi \right) \partial_\mu \varphi\, .$$
Such an extra interaction gives zero only if the external line spinors obey the
Dirac equation. To see that requires a derivation having some similarity to the
work needed when demonstrating gauge invariance of the diagrams in quantum
electrodynamics.

The transformation that we wish to perform in the case of quantum electrodynamics
may be written as the product of two transformations:
$$ \psi \rightarrow \left( -\gamma \partial + m\right)
 \left( 1 - {1 \over -\gamma \partial + m}\, ie\gamma^\mu A_\mu \right) \psi\, .$$
The first transformation is the one we discussed above. The second is now in a form
that shows clearly that we are dealing with a non-local transformation. The structure
is very suggestive: a fermion propagator times the usual Dirac e.m. interaction.
It will give rise to Faddeev-Popov ghost loops. The Jacobian of the transformation
can be established without any problem; alternatively one can use combinatorial
methods to deduce the precise form of the ghost loops (see ref. [6], in particular
section 10.4). That latter approach has the advantage that it results in a detailed
understanding of the structure of the loops, and the associated sign. The result
is very simple: the Faddeev-Popov loops are precisely as those of the Dirac theory,
however with a plus sign. That plus sign can be understood as a combination of the
minus sign for a fermion loop combined with a minus sign for the Faddeev-Popov ghost
loop. This may be re-arranged in an elegant manner. Consider as an example any
diagram with one closed loop in the Dirac theory. Such a diagram has a minus sign.
In the new theory we will have a loop generated by the new rules (which must be
provided with the usual minus sign for a fermion loop), plus a loop
precisely equal to the original loop of the Dirac theory, however now with a plus
sign. Thus, $-$ Dirac loop = $-$ New loop $+$ Dirac loop. Or, we can simply take the
Faddeev-Popov loops into account by giving the new loops a factor \smn12.

Clearly, all this is quite complicated and not very transparent. However, the
results of this section can be reproduced quite simply for the diagrams of the
theory, without any considerations concerning Faddeev-Popov ghosts. In that way
we can verify the rules for the new theory without leaving any lingering doubts.

\section{Diagrams}%
The Feynman rules for the Dirac theory are as usual:
\smallskip
\setbox0=\vbox{\hsize=80mm\noindent
\noindent Electron propagator\hskip 6mm $\displaystyle{-i\gamma p +m \over p^2+m^2-i\epsilon}$}

\line{\hskip 10mm\raise 2mm\box0\hskip 10mm\epsffile{mom_elpr.eps}}
\smallskip
\setbox0=\vbox{\hsize=80mm\noindent
\noindent  Photon propagator\hskip 10mm
 $\displaystyle{\delta_{\mu\nu} \over k^2-i\epsilon}$}

\line{\hskip 10mm\raise 2mm\box0\hskip 10mm\epsffile{mom_phop.eps}}

\setbox0=\vbox{\hsize=80mm\noindent
\noindent  Vertex \hskip 30mm  $\displaystyle{-ie\gamma^\mu}\, .$}

\line{\hskip 10mm\raise 5mm\box0\hskip 8.5mm\epsffile{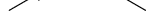}}

\noindent The Feynman rules for the `two-component' theory are:
\smallskip
\setbox0=\vbox{\hsize=80mm\noindent
\noindent Electron propagator\hskip 10mm $\displaystyle{1 \over p^2+m^2-i\epsilon}$}

\line{\hskip 10mm\raise 2mm\box0\hskip 10mm\epsffile{mom_elpr.eps}}
\smallskip
\setbox0=\vbox{\hsize=80mm\noindent
\noindent  Photon propagator\hskip 14mm
 $\displaystyle{\delta_{\mu\nu} \over k^2-i\epsilon}$}

\line{\hskip 10mm\raise 2mm\box0\hskip 10mm\epsffile{mom_phop.eps}}

\smallskip
\noindent and furthermore a three-point and a four-point vertex:

\setbox0=\vbox{\hsize=80mm\noindent
$\displaystyle{ -ie V_\mu(p,k)= ie(2ip_\mu+ik_\mu+2i\sigma^{\nu\mu}k_\nu)}$}
\setbox1=\vbox{\hsize=35mm\noindent
$\displaystyle{ - 2 e^2 \delta_{\mu\nu}}\, .$}

\line{\hskip 10mm\raise 5mm\box0\hskip 8mm\epsffile{mom_ephv.eps}}

\line{\hskip 55mm\raise 5mm\box1\hskip 8.5mm\epsffile{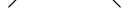} }

As will be made clear, every string of gamma matrices terminated by spinors must
be provided with a factor $1/2m$. Those spinors are the usual ones, precisely as
in the Dirac theory. Closed fermion loops have a minus sign and furthermore a
factor \smn12 must be provided.

The convention for the sign of the momenta in the three-point vertex is as usual,
i.e. the momenta $p$ and $k$ are in going.

It should be noted that the propagator must still be drawn with an arrow,
which is of importance in the vertices to which the propagators connect.

There are two basic identities needed to perform the change-over from the Dirac
theory to the `two-component' theory on the diagram level.
Consider the vertex of the Dirac theory as occurring in some diagram,
together with an electron propagator attached to the incoming electron line
at that vertex. The momenta of the incoming electron and the photon will be
called $p$ and $k$ respectively. There will be a factor $-i\gamma p + m$
associated with the propagator of the line with momentum $p$, and we want
to move this factor to the other line through the gamma matrix in the vertex.
The momentum of that line (the outgoing electron) will be denoted by $p_1 = p+k$:
$$ \eqalignno{\gamma^\mu (-i\gamma p +m) &= (i\gamma p +m)\gamma^\mu -2i p_\mu\cr
 &= (i\gamma p_1 +m)\gamma^\mu -i\gamma k - 2ip_\mu\cr
 &= (i\gamma p_1 +m) \gamma^\mu
- \smn{i}2 (\gamma^\nu \gamma^\mu - \gamma^\mu \gamma_\mu) k_\nu
 - ip_{1\mu} - ip_\mu\cr
 &= (i\gamma p_1 +m) \gamma^\mu
- 2i \sigma^{\nu\mu}k_\nu
 - ip_{1\mu} - ip_\mu\,. }$$
The reader may recognize the three-point vertex of the `two-component'
theory in the last three terms of this expression. If we abbreviate
 $$ V_\mu(p,k) =- 2i \sigma^{\nu\mu}k_\nu - i(p+k)_\mu - ip_\mu $$
then our first basic identity is
$$ \gamma^\mu (-i\gamma p + m) = (i\gamma p_1 + m)\gamma^\mu + V_\mu(p,k)\, .$$
We now use this identity on a string such as encountered in diagrams. In a diagram,
following a fermion line, we encounter in the Dirac theory vertex factors $\gamma^\mu$
separated by propagators. Let us denote such a string, beginning and ending with
a vertex (a $\gamma$ matrix) by {\bf S}; it is of the form:
$$ {\bf S} =\gamma^{\mu_n} {-i\gamma p_n + m \over p_n^2+m^2-i\epsilon}
  \gamma^{\mu_{n-1}}\ldots
 {-i\gamma p_2 + m \over p_2^2+m^2-i\epsilon}
  \gamma^{\mu_1} {-i\gamma p_1 + m \over p_1^2+m^2-i\epsilon}
  \gamma^{\mu}\, .$$

\setbox0=\vbox{\hsize=60mm\noindent
The figure shows the type of diagram substructure corresponding to such an
expression. We have also indicated momenta as used below.}

\line{\box0\hfil\epsffile{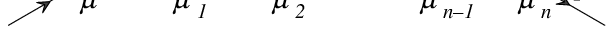}}

Of course, one must for a given process consider all possible diagrams; of
relevance here are the strings obtained by permuting the photon lines. In
order to keep things transparent we will not indicate that explicitly, but
discuss the consequence of that where relevant.

Consider such a string multiplied on the right with the energy projection
operator:
$$ {\bf S} (-i\gamma p + m)\, .$$
Here $p$ is the initial momentum, as shown in the figure above. Now use the
basic identity once, involving $\gamma^\mu$. Writing only the last part of
the string we get:
$$\eqalignno{{\bf S}(-i\gamma p + m) =& \ldots {-i\gamma p_2 + m \over p_2^2+m^2-i\epsilon}
 \gamma^{\mu_1} {-i\gamma p_1 + m \over p_1^2+m^2-i\epsilon}
  \left((i\gamma p_1 + m) \gamma^\mu + V_\mu(p,k)\right)\cr
 =& \ldots {-i\gamma p_2 + m \over p_2^2+m^2-i\epsilon}\gamma^{\mu_1}\gamma^\mu \cr
 & + \ldots  {-i\gamma p_2 + m \over p_2^2+m^2-i\epsilon}
 \gamma^{\mu_1} {-i\gamma p_1 + m \over p_1^2+m^2-i\epsilon} V_\mu (p,k)\, .}$$
The first term involves the product of two $\gamma$'s, namely
$\gamma^{\mu_1}\gamma^\mu$. It is here that it is important to include diagrams
with the photon lines permuted. One of them will give precisely the same string
as above, but with the indices $\mu_1$ and $\mu$ interchanged. We therefore,
without explicitly including the diagrams with permuted photon lines, replace
this product by $\delta_{\mu_1\mu}$. Together with the permuted diagram, and
including the factors $-ie$ of the Dirac vertices we so obtain the four-point
vertex of the `two-component' theory.

The steps taken so far may be repeated, in the first term using the energy
projection operator $-i\gamma p_2+m$, and in the second term $-i\gamma p_1+m$.
In this way, moving the energy projection operator to the left, we arrive at
the final result, to be called the string equation:
$$ {\bf S}(-i\gamma p + m) = {\bf S_2} +
 (i\gamma p' + m)\gamma^{\mu_n} {1 \over p^2_n + m^2}{\bf S'_2}\, .$$
where ${\bf S_2}$ and ${\bf S'_2}$ denote strings of propagators and vertices of
the `two-component' theory, always beginning and ending with a three- or four-point
vertex. More precisely, they are sums of strings, with the three- and four-point
vertices distributed in all possible ways.

We are ready to consider the $S$-matrix in detail. There are two types of
fermion lines, namely through-going lines terminated by spinors on both sides
and closed loops. First a fermion line in the Dirac theory with spinors at the
beginning and end. The spinors obey the Dirac equation:
$$(i\gamma p + m)u(p)=0 \quad\quad{\rm and}\quad\quad
\overline{u}(p')(i\gamma p' + m)=0\, .$$
Consider an expression of the form
$$ \overline{u}(p')\, {\bf S}\, u(p)\, .$$
Since $u(p)$ obeys the Dirac equation we may write:
$$ u(p) = {1 \over 2m} (-i\gamma p + m) u(p)\, .$$
Using our string equation for the combination ${\bf S}(-i\gamma p+m)$ we find, using
the Dirac equation for $\overline{u}(p')$:
$$  \overline{u}(p')\, {\bf S}\, u(p) =
 {1 \over 2m} \overline{u}(p')\, {\bf S}\,(-i\gamma p + m) u(p)
 = {1 \over 2m}  \overline{u}(p')\, {\bf S_2}\, u(p)\, .$$
Thus we find for those through-going lines the strings corresponding to the
`two-component' theory, including the factor $1/2m$.

Now closed loops. That is ever so slightly more complicated. Consider a closed loop
and choose a propagator. Exhibiting this propagator explicitly the closed loop takes
the form:
$$ {\rm Tr}\left[ {\bf S}\, { -i\gamma p + m \over p^2+m^2-i\epsilon} \right]\, .$$
Using the identity
$$ (-i\gamma p + m)^2 = 2m (-i\gamma p + m) - (p^2 + m^2)$$
we substitute
$$ {-i\gamma p + m\over p^2+m^2-i\epsilon} =
 {1 \over 2m}(-i\gamma p + m) {-i\gamma p + m \over p^2+m^2-i\epsilon}
 + {1 \over 2m} \, .$$
Let us concentrate on the first term. We have:
$$\eqalignno{ {1 \over 2m}& {\rm Tr}\left[
 {\bf S}\, (-i\gamma p+ m) { -i\gamma p + m \over p^2+m^2-i\epsilon}\right]
 = {1 \over 2m} {\rm Tr}\left[ {\bf S_2}\, {-i\gamma p + m \over p^2+m^2-i\epsilon}\right]\cr
 &\quad\quad\quad\quad
 + {1 \over 2m} {\rm Tr}\left[{\bf S_2'}\, {-i\gamma p + m \over p^2+m^2-i\epsilon}
 (i\gamma p + m) \gamma^{\mu_n} {1 \over p_n^2+m^2-i\epsilon}\right]\cr
 =&\ {1 \over 2m} {\rm Tr}\left[ {\bf S_2}\,
 {-i\gamma p + m \over p^2+m^2-i\epsilon}\right]
 + {1 \over 2m} {\rm Tr}\left[{\bf S_2'}\, 
 \gamma^{\mu_n} {1 \over p_n^2+m^2-i\epsilon}\right]\cr
 \, .}$$
At this point it must be remembered that the strings ${\bf S_2}$ and ${\bf S_2'}$
involve an even number of $\gamma$'s only. Since the trace of an odd number of $\gamma$'s
is zero the  second term is zero and from the first term only the part with $m$ in
the numerator survives. We thus find:
$$ {\rm Tr}\left[ {\bf S}\, { -i\gamma p + m \over p^2+m^2-i\epsilon} \right]
 = {1 \over 2} {\rm Tr}\left[{\bf S_2} {1 \over p^2+m^2}\right]
 + {1 \over 2m} {\rm Tr}\left[ {\bf S} \right]\, .$$
Remember that ${\bf S_2}$ starts and ends with a vertex. All together this is indeed
a trace as encountered in the `two-component' theory, and we also see the factor two,
understood as due to a Faddeev-Popov ghost. However, note that the first term has
an explicit propagator at the location of the propagator that we started from, and
we thus still miss a piece, namely a `two-component' type diagram with a four-vertex
at that location. That however, is precisely what the last term provides, but to
see that we must still do some work. Peeling off one vertex and
one propagator from the Dirac string ${\bf S}$ we find:
$$\eqalignno{ {1 \over 2m}{\rm Tr}\left[ {\bf S} \right] &=
 {1 \over 2m} {\rm Tr}\left[ {\bf S'} {-i \gamma p_1 + m\over p_1^2+m^2-i\epsilon}
 \gamma^\mu\right]\cr
 & = {1 \over 2m}
 {\rm Tr}\left[ {\bf S_2'}\, {1 \over p_1^2+m^2} \gamma^\mu\right]\cr
 &\quad\quad + {1 \over 2m} {\rm Tr}\left[ (i\gamma p' + m) \gamma^{\mu_n}
  { 1 \over p_n^2+m^2}
  {\bf S''_2}\,{1 \over p_1^2+m^2}\gamma^\mu\right]\cr
 & = {1 \over 2}
 {\rm Tr}\left[ \gamma^{\mu_n} { 1 \over p_n^2+m^2}
  {\bf S''_2}\,{1 \over p_1^2+m^2}\gamma^\mu\right]\cr
 & ={1 \over 2} {\rm Tr}\left[ {\bf S''_2}\,{1 \over p_1^2+m^2}\gamma^\mu
  \gamma^{\mu_n} { 1 \over p_n^2+m^2}\right] \, .} $$
We used that the trace over a product of an odd number of $\gamma$'s is zero. Only
the part containing $m$ survives. We note the product
$\gamma^\mu\gamma^{\mu_n}$, and (remembering the
argument about including diagrams with permuted photon lines) we recognize
a four-point vertex  with its surrounding propagators. Note that the propagator
of momentum $p$ has disappeared. In other words, this term generates a diagram
with a four-point vertex at the location of the propagator that we started from.
That is indeed the missing piece. Taking all together we thus find:
$$ {\rm Tr}\left[ {\bf S} \right] =
 {1 \over 2} {\rm Tr}\left[ {\bf S_2} \right]\, .$$

\section{The two-component theory}%
The transition from the `two-component' to the true two-component theory
relies on a remarkable identity. Inserting $\gamma^5$ in either a through going
fermion line or a loop in the `two-component' theory makes those diagrams zero.
Consider a through going line, and insert $\gamma^5$ in front of the spinor $u$:
$$ \overline u(p'){\bf S_2}\, \gamma^5 u(p)\, .$$
Using our string identity backwards this is equal to:
$$\eqalignno{&\left(\overline u(p') {\bf S}\, (-i\gamma p + m)\gamma^5 u(p)\right)\cr
 &\quad\quad\quad - \left(\overline u(p') (i\gamma p' + m)\gamma^{\mu_n}
 {1 \over p_2^2+m^2-i\epsilon}
 {\bf S'_2}\, \gamma^5 u(p)\right)=0\, .}$$
Using $(-i\gamma p + m)\gamma^5=\gamma^5(i\gamma p + m)$ in the first term we see that
it is zero because $u(p)$ obeys the Dirac equation; similarly the second term is
zero for the same reason with respect to $\overline u(p')$.

For loops the reasoning is slightly more complex due to the complications relating
to the four-point vertex. Let us consider a closed loop (or rather the sum of
closed loops of a given order with all possible distributions of the three- and
four-point vertices), and select a propagator. There will be a diagram were that
propagator is missing, having a four-point vertex at that spot. If again ${\bf S_2}$
(and ${\bf S'_2}$) denotes strings with a vertex at beginning and end we have,
introducing a $\gamma^5$ as well, an expression of the form:
$$ {\rm Tr}\left[ {\bf S_2}\, {1 \over p^2+m^2} \gamma^5 \right]
  + {\rm Tr}\left[ {\bf S'_2}\,{1 \over p_1^2+m^2}\gamma^\mu
  \gamma^{\mu_n} { 1 \over p_n^2+m^2}\gamma^5\right] \, . $$
In the last term we move $\gamma^5$ one place to the left, getting a crucial minus
sign:
$$ {1 \over 2}{\rm Tr}\left[ {\bf S_2}\, {1 \over p^2+m^2} \gamma^5 \right]
  - {1 \over 2}{\rm Tr}\left[ {\bf S'_2}\,{1 \over p_1^2+m^2}\gamma^\mu\gamma^5
  \gamma^{\mu_n} { 1 \over p_n^2+m^2}\right] \, . $$
Now doing the reverse work, going back to Dirac strings we find as result:
$$ {1 \over 2m}{\rm Tr}\left[ {\bf S}\, (-i\gamma p + m)\gamma^5
 {-i\gamma p + m \over p^2+m^2-i\epsilon}\right]
 - {1 \over 2m}{\rm Tr}\left[ {\bf S}\, \gamma^5\right]\, .$$
The reader will have no difficulty verifying that the two terms cancel, and we
find again zero.

The transition to the true two-component theory is now very simple: we simply
insert a helicity projection operator \smn{1+\gamma^5}2 in loops as well as
in through going lines. That takes care of the factor \smn12 associated with
closed loops; our spinor normalization must now include a factor $1/\sqrt{m}$.
Since the square of such a projection operator is that operator itself one
may actually put one at every vertex. Note that we have no need for
$\smn{1-\gamma^5}2$. In a representation were $\gamma^5$ is diagonal (see
appendix) one obtains the two-component theory in all detail, because
a helicity projection operator applied to a four-component spinor makes two of
its components equal to zero.

Doing calculations in the two-component theory one needs the expression for the
sum over spins of the product $u(p)\overline u(p)$. Our spinors are still the
usual Dirac spinors, so that is as usual:
$$ \sum_{\rm spins} u(p)\overline u(p) = { 1\over 2E} (-i\gamma p + m)\, .$$
This expression will have factors $\smn{1+\gamma^5}2$ on both sides, and therefore
the piece $-i\gamma p$ will not contribute. Including a factor $1/m$ in the spinor
normalization we find a very simple result, namely just $1/2E$. In the
literature the factor $1/2E$ is often put somewhere else, and with that convention
the result is simply one (or rather the two-by-two unit matrix $\sigma^0$), as stated
for example by Brown [2].

\section{Anomalous magnetic moment: Compton scattering}%
Having understood the details of the `two-component' theory it is now easy
to study the consequences of an anomalous magnetic moment of the electron.
For this purpose we provide the magnetic term in the interaction (in the
`two-component' formulation) with a factor $\kappa$. If $\kappa \neq 1$
there is an anomalous magnetic moment; here however we do not intend to study
an anomalous moment, but rather use $\kappa$ to label the terms due
to the magnetic part of the interaction. The Feynman rules are as shown before,
except that we provide the term $\sigma^{\mu\nu}k_\nu$ with a factor $\kappa$.

We consider here a very simple case, namely Compton scattering. The known
expression for that process is remarkable for its simplicity, and it might
be interesting to separate out the electric from the magnetic part.
This process was also considered by Brown [2], although with
specially chosen polarization vectors and no explicit separation of
the magnetic part.

\setbox0=\vbox{\hsize=65mm
In lowest non-vanishing order there are two diagrams contributing to the
amplitude, see figure. As variables we use the two dot-products $pk$ and\break
\vskip -\baselineskip}
\vskip 1mm
\line{\box0\hfil\epsffile{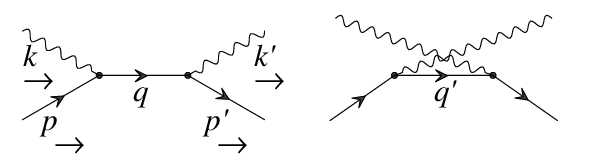}}

\noindent $pk'$ for which we substitute $-m\omega$ and $-m\omega'$. In the rest
system of the initial electron, $\omega$ and $\omega'$ are then the
energies of the initial and final photon respectively. The result is
(we have not bothered to work out the overall factor)
$$ \eqalignno{\sum_{\rm el.\ spins}|{\cal A}|^2 \propto\ &
 {e^4 \over \omega^2\omega'^2}\Biggl[
 \kappa^4\left(-2\omega\omega'(\omega-\omega')^2
 +8{\omega^2\omega'^2 \over m}(\omega-\omega') \right)\cr
 &+ \kappa^3
\left(8\omega\omega'(\omega-\omega')^2
  -16{\omega^2\omega'^2 \over m}(\omega-\omega')\right)\cr
 &+\kappa^2\left( 2m(\omega-\omega')^3
 -10\omega\omega'(\omega-\omega')^2 + 12{\omega^2\omega'^2 \over m}(\omega-\omega')\right)\cr
 &+ \kappa\left(-4m(\omega-\omega')^3
 + 12\omega\omega'(\omega-\omega')^2 - 8{\omega^2\omega'^2 \over m}(\omega-\omega') \right)\cr
 &+ 4\omega\omega'(\omega^2+\omega'^2) + 2m(\omega-\omega')^3
 +(4m^2-8\omega\omega')(\omega-\omega')^2\cr
 &+ \left(4{\omega^2\omega'^2 \over m}
  - 8m\omega\omega'\right) (\omega-\omega') \Biggr]
}$$

For $\kappa=1$ the result reduces to the standard result:
$$\sum_{\rm el.\ spins}|{\cal A}|^2 \propto\ 
 {e^4 \over \omega^2\omega'^2}\left[4\omega\omega'(\omega^2+\omega'^2)
 +4m^2(\omega-\omega')^2-8m\omega\omega'(\omega-\omega')\right]$$

The final expression is surprisingly simple, but that simplicity is lost if the magnetic
moment terms are kept separately. It is hard to see the physics of it. While one
might expect that magnetic moment terms would contain higher powers of the photon
momenta, this turns out not to be the case. The electron propagators are inversely
proportional to these same momenta, and nothing special survives. The physics
picture remains complicated.

\section{Anomalous magnetic moment, renormalizabilty and unitarity}
Finally we will discuss briefly the more theoretical aspects of an anomalous
magnetic moment. The Lagrangian in the `two-component' formulation including an
anomalous moment is
$$\eqalignno{ {\cal L}^{\rm anom}_2
 =&\ - \smn12 \partial_\mu A_\nu \partial_\mu A_\nu
 -(\partial_\mu -ieA_\mu) \overline \psi (\partial_\mu +ie A_\mu)\psi
 - m^2 \overline \psi \psi\cr
 &\ + 2ie(1+\lambda)
 \left( \overline \psi \sigma^{\mu\nu} \psi\right) \partial_\mu A_\nu\, .}$$
This Lagrangian appears to be of a renormalizable type, despite the addition
of an anomalous magnetic moment.
It is simple to see how to modify the Dirac formulation in order to generate
the extra term:
$$\eqalignno{{\cal L}^{\rm anom}_{\rm D} &=
 - \smn12 \partial_\mu A_\nu \partial_\mu A_\nu
 - \overline \psi (\gamma^\mu D_\mu + m)\psi\cr
 &\quad\quad\quad+ 2ie\lambda \Bigl( \overline \psi \sigma^{\mu\nu}
 {1 \over - \gamma^\alpha D_\alpha + m} \psi\Bigr) \partial_\mu A_\nu\, .}$$
This is a non-local interaction that will in general violate unitarity. The
extra term can be expanded:
$$ {1 \over -\gamma \partial + m - ie\gamma^\alpha A_\alpha}
 = {1 \over -\gamma \partial + m}\left( 1 +
  {1\over -\gamma\partial + m}\,ie\gamma^\alpha A_\alpha  + \ldots\right)\, .
$$
It is clear that this term interferes with our transformation, shifting the
factor $-\gamma \partial + m$ from the propagator to the vertex. We therefore
loose unitarity, even if the interaction appears local in the `two-component
theory. The theory is not acceptable for $\lambda \neq 0$.

It is interesting though that an anomalous magnetic moment term in the two
component theory is not equivalent to a Pauli term. There is another observation
that can be made here. In the `two-component' formulation with an anomalous moment
the string equation will not be valid anymore. Inserting a $\gamma^5$ will not
give zero. Therefore the true two-component Lagrangian (with only
$\overline \psi_R \psi_L$ terms) will no more produce a unitary $S$-matrix,
because the hermitean conjugate of that Lagrangian (with $\overline \psi_L \psi_R$
terms) produces a different $S$-matrix.

\section{Conclusions}%
The two-component theory provides for an interesting exercise in field theory
and diagrammatic methods. Also, it neatly separates out the electric and magnetic
parts of the interaction of photons with electrons, which may be of interest
in certain situations. Unitarity however is not automatic in that formalism, and
needs separate verification. An anomalous magnetic moment in the two-component
theory violates unitarity, as also noted explicitly in ref. [3].

The original motivation to re-introduce the two-component theory, namely as
a formalism appropriate to the study of the theory of weak interactions, has
somehow been obscured. We may ask whether the transformation introduced in
this paper may with advantage be applied to the Standard Model. We have not
investigated that.

\section{Acknowledgments}%
The author is very grateful to Prof. V. Telegdi, who provided the inspiration
to study the present subject. Furthermore the author is indebted to
Profs. R. Stora and J. van der Bij for their critical remarks and
helpful comments.

\appendix{Spinor representations}%
The most common representation has diagonal $\gamma^4$:
$$\gamma^1 = \left[\matrix{0&0&0&-i\cr 0&0&-i&0\cr 0&i&0&0\cr
i&0&0&0\cr}\right] \qquad \gamma^2 = \left[\matrix{0&0&0&-1\cr
0&0&1&0\cr 0&1&0&0\cr 
-1&0&0&0\cr}\right]$$
$$\gamma^3 = \left[\matrix{0&0&-i&0\cr 0&0&0&i\cr i&0&0&0\cr
0&-i&0&0\cr}\right] \qquad \gamma^4 = \left[\matrix{1&0&0&0\cr
0&1&0&0\cr 0&0&-1&0\cr 0&0&0&-1\cr}\right]$$
$$\gamma^5 =
\left[\matrix{0&0&-1&0\cr 0&0&0&-1\cr -1&0&0&0\cr 0&-1&0&0\cr}\right]$$
where $\gamma^5 = \gamma^1 \gamma^2 \gamma^3 \gamma^4.$ Using the $\sigma$
matrices including the identity $\sigma^0$:
$$\sigma^0=\left(\matrix{1&0\cr 0&1\cr}\right)\quad
\sigma^1=\left(\matrix{0&1\cr 1&0\cr}\right)\quad
 \sigma^2=\left(\matrix{0&-i\cr i&0\cr}\right)\quad
 \sigma^3=\left(\matrix{1&0\cr 0&-1\cr}\right)$$
this can also written as
$$\gamma^j = \left(\matrix{0&-i\sigma^j\cr i\sigma^j&0}\right)\quad\quad
 \gamma^4 = \left(\matrix{\sigma^0&0\cr 0&-\sigma^0}\right)\quad\quad
 \gamma^5 = \left(\matrix{0&-\sigma^0\cr -\sigma^0&0}\right)\, .$$
In this representation the spinors are ($E=\sqrt{\vec p^{\,2}+m^2}$):

$$\vbox{\offinterlineskip\halign{
\hfil$\displaystyle{#}$\quad\hfil
&\strut\vrule\ \hfil$\displaystyle{#}$\ \hfil
&\vrule\ \hfil$\displaystyle{#}$\ \hfil
&\vrule\ \hfil$\displaystyle{#}$\ \hfil
&\vrule\ \hfil$\displaystyle{#}$\ \hfil
&\vrule$\displaystyle{#}$\quad\cr
&\multispan4\hrulefill\cr
\phantom{a \over a}&u^1(p)\uparrow&u^2(p)\downarrow
&u^3(p)\downarrow&u^4(p)\uparrow&\cr
&\multispan4\hrulefill\cr
\phantom{E \over E}&1&0& {-p_3 \over m+E}
& {p_1-ip_2 \over m+E}&\cr
\times\sqrt{m+E \over 2E}&0&1& -{p_1+ip_2 \over m+E}
& {-p_3 \over m+E}&\cr
\phantom{E \over E_j}&{p_3 \over m+E}
& {p_1-ip_2 \over m+E} &-1&0&\cr
\phantom{E \over E_j}&{p_1+ip_2 \over m+E}
& {-p_3 \over m+E} &0&1&\cr
&\multispan4\hrulefill\cr}}$$
\smallskip
The first two columns are the particle spinors; they are solutions
of the equation $(i\gamma p + m)u(p)=0$. The last two columns are the
anti-particle spinors, solutions of the equation $(-i\gamma p + m)u(p)=0$.
One may verify:
$$ u^{1,2}(p) = {-i\gamma p + m \over \sqrt{2E(m+E)}}\ \ v^{1,2}\quad\quad
 u^{3,4}(p) = {i\gamma p + m \over \sqrt{2E(m+E)}}\ \ v^{3,4}$$
where the $v^j$ are equal to the $u^j$ for zero momentum (and thus also $E=m$):
$v^j = u^j(0)$.

Concerning the sum over spins one has:
$$\eqalignno{\sum^2_{j=1} u^j \left(\vec p\right) \bar
u^j \left(\vec p\right) &= {1\over 2E}\left(-i\gamma p
+m\right)\quad\quad{\rm and}\quad\quad \sum^2_{j=1} v^j \bar v^j = \smn12 (\gamma^4+1)\cr
\sum^4_{j=3} u^j \left(\vec p\right) \bar
u^j \left(\vec p\right) &= {1\over 2E}\left(-i\gamma p
- m\right)
\quad\quad{\rm and}\quad\quad \sum^2_{j=1} v^j \bar v^j = \smn12 (\gamma^4-1).\cr}$$
The relation between the spin sums for the $u$ and $v$ can be made explicit:
$$\eqalignno{\sum^2_{j=1} u^j \left(\vec p\right) \bar
u^j  &= {-i\gamma p + m \over \sqrt{2E(m+E)}}\,
 \smn12 (\gamma^4+1)\,{-i\gamma p + m \over \sqrt{2E(m+E)}}\cr
 &={-i\gamma p + m \over 4E(m+E)}\left(2m - 2ip_4\right)={-i\gamma p + m \over 2E}\, .
}$$

A representation in which $\gamma^5$ is diagonal can be obtained by means
of the transformation
$$ T = {1 \over \sqrt{2}}
 \left(\matrix{\sigma^0&-\sigma^0\cr\sigma^0&\sigma^0\cr}\right)\,,
 \quad T^{-1} ={1 \over \sqrt{2}}
 \left(\matrix{\sigma^0&\sigma^0\cr-\sigma^0&\sigma^0\cr}\right)\, .$$
The $\gamma^j$, $j=1,2,3$ are invariant under this transformation.
For $T\gamma^4 T^{-1}$ and $T\gamma^5 T^{-1}$ one finds:

$$ \gamma^4 = \left(\matrix{0&\sigma^0\cr \sigma^0&0}\right)\quad\quad
 \gamma^5 = \left(\matrix{\sigma^0&0\cr 0&-\sigma^0}\right)\, .$$
Thus simply $\gamma^5 \rightarrow \gamma^4$, $\gamma^4 \rightarrow -\gamma^5$.
The combination \smn{1+\gamma^5}2 when applied to a spinor projects out the
upper two components. The transformed spinors $u^j(p)\rightarrow Tu^j(p)$ are:

$$\vbox{\offinterlineskip\halign{
\hfil$\displaystyle{#}$\quad\hfil
&\strut\vrule\ \hfil$\displaystyle{#}$\ \hfil
&\vrule\ \hfil$\displaystyle{#}$\ \hfil
&\vrule\ \hfil$\displaystyle{#}$\ \hfil
&\vrule\ \hfil$\displaystyle{#}$\ \hfil
&\vrule$\displaystyle{#}$\quad\cr
&\multispan4\hrulefill\cr
\phantom{a \over a}&u^1(p)\uparrow&u^2(p)\downarrow
&u^3(p)\downarrow&u^4(p)\uparrow&\cr
&\multispan4\hrulefill\cr
\phantom{E \over E}&1-{p_3 \over m+E}&-{p_1-ip_2 \over m+E}&1- {p_3 \over m+E}
& {p_1-ip_2 \over m+E}&\cr
\times\sqrt{m+E \over 4E}&-{p_1+ip_2 \over m+E}&1+{p_3 \over m+E}& -{p_1+ip_2 \over m+E}
& -1-{p_3 \over m+E}&\cr
\phantom{E \over E_j}&1+{p_3 \over m+E}
& {p_1-ip_2 \over m+E} &-1- {p_3 \over m+E}&{p_1-ip_2 \over m+E}&\cr
\phantom{E \over E_j}&{p_1+ip_2 \over m+E}
& 1-{p_3 \over m+E} &-{p_1+ip_2 \over m+E}&1-{p_3 \over m+E}&\cr
&\multispan4\hrulefill\cr}}$$
\smallskip
\noindent and the $v\rightarrow Tv$:
$$\vbox{\offinterlineskip\halign{
\hfil$\displaystyle{#}$\quad\hfil
&\strut\vrule\ \hfil$\displaystyle{#}$\ \hfil
&\vrule\ \hfil$\displaystyle{#}$\ \hfil
&\vrule\ \hfil$\displaystyle{#}$\ \hfil
&\vrule\ \hfil$\displaystyle{#}$\ \hfil
&\vrule$\displaystyle{#}$\quad\cr
&\multispan4\hrulefill\cr
\phantom{a \over a}&v^1\uparrow&v^2\downarrow
&v^3\downarrow&v^4\uparrow&\cr
&\multispan4\hrulefill\cr
\phantom{E \over E}&1&0&1&0&\cr
\times{1 \over \sqrt{2}}&0&1&0& -1&\cr
\phantom{E \over E_j}&1&0&-1&0&\cr
\phantom{E \over E_j}&0& 1&0&1&\cr
&\multispan4\hrulefill\cr}}$$

\references
\halign{\hfil[#] &\vtop{\parindent=0pt\hsize=119mm
\hangindent0pt\strut#\strut}\cr
1& R.P. Feynman and M. Gell-Mann, {\it Phys. Rev.} {\bf109} (1958) 193.\cr
2& L.M. Brown, {\it Phys. Rev.} {\bf 111} (1958) 957.\cr
3& W.R. Theis, {\it Forts. der Physik} {\bf 7} (1959) 559.\cr
4& M. Tonin, {\it Il Nuovo Cim.} {\bf 14} (1959) 1108.\cr
5& M. Veltman, {\it Diagrammatica}, Cambridge University Press, 1994.\cr
6& G. 't Hooft and M. Veltman, {\it Diagrammar}, CERN yellow report 73-9, 1973.\cr
}

\end